\documentclass{emulateapj}
\newcommand{\cmden}{\mbox{ cm$^{-3}$}}
\newcommand{\cmsq}{\mbox{ cm$^{2}$}}
\newcommand{\Mpc}{\mbox{ Mpc}}

\newcommand{\erg}{\mbox{ erg}}

\newcommand{\eV}{\mbox{ eV}}
\newcommand{\kel}{\mbox{ K}}

\newcommand{\ergsec}{\mbox{ erg s$^{-1}$}}

\newcommand{\yr}{\mbox{ yr}}

\newcommand{\secinv}{\mbox{ s$^{-1}$}}

\newcommand{\Msun}{\mbox{ M$_\odot$}}

\newcommand{\Lsun}{\mbox{ L$_\odot$}}

\newcommand{\hunits}{\mbox{ km s$^{-1}$ Mpc$^{-1}$}}

\newcommand{\lya}{Ly$\alpha$ }

\newcommand{\bq}{\begin{equation}}
\newcommand{\eq}{\end{equation}}
\newcommand{\bqa}{\begin{eqnarray}}
\newcommand{\eqa}{\end{eqnarray}}
\def\VEV#1{\left\langle #1\right\rangle} 

\begin{document}

\title{Fluctuations in the Ionizing Background During and After Helium Reionization}

\author{Steven R.  Furlanetto}

\affil{Department of Physics and Astronomy, University of California, Los Angeles, CA 90095, USA; email: sfurlane@astro.ucla.edu}

\begin{abstract}
The radiation background above the ionization edge of \ion{He}{2} varies strongly during and after helium reionization, because the attenuation length of such photons is relatively short ($\la 40 \Mpc$) and because the ionizing sources (quasars) are rare. Here we construct analytic and Monte Carlo models to examine these fluctuations, including, for the first time, those during the reionization era itself.  In agreement with detailed numerical simulations, our analytic model for the post-reionization Universe predicts order-of-magnitude fluctuations in the \ion{He}{2} ionization rate $\Gamma$.  Observations of the hardness ratio \ion{He}{2}/\ion{H}{1} show somewhat larger fluctuations, which may be due to more complicated radiative transfer effects.  During reionization, the fluctuations are even stronger.  In contrast to hydrogen reionization, our model predicts that regions with strong \ion{He}{2} \lya forest transmission should be reasonably common even during the beginning stages of reionization, because of strong illumination from nearby bright quasars.  Partly because of this, the mean ionizing background does not evolve strongly during and after helium reionization; it is roughly proportional to the filling fraction of \ion{He}{3} regions.  On the other hand, regions full of \ion{He}{2} and also ``fossil" ionized regions that contain no (or few) active sources appear as strong IGM absorbers.  Their presence exaggerates the evolution of the hardness ratio, making it evolve more strongly than naively expected during the reionization era.
\end{abstract}
  

\section{Introduction} \label{intro}

To the vast majority of baryonic matter in the Universe, the most important radiation field is the metagalactic ionizing background.  As a result, a great deal of effort has gone into measuring its properties.  During most of cosmic history, the background evolved relatively slowly, but there were two major exceptions:  the reionization of hydrogen (at $z \ga 6$) and of helium (with double ionization occurring at $z \sim 3$).  During each of these episodes, the intergalactic medium (IGM) underwent a phase transition and the Universe became (mostly) transparent to the relevant ionizing photons, allowing the high-energy radiation field to grow rapidly.

Recently, these events have received a great deal of attention in both the observational and theoretical communities (see, e.g., reviews by \citealt{barkana01, ciardi05-rev, fan06-review, furl06-review} about hydrogen reionization), and our understanding of the evolution of the ionizing radiation field has become increasingly sophisticated.  However, most such studies still treat the ionizing background as spatially \emph{uniform}.

During reionization itself, this assumption is obviously wrong, because some regions are exposed to strong ionizing radiation while others remain neutral with no local illumination.  Moreover, even within the ionized regions, differences in the effective ``horizon" to which ionizing sources can be seen induce large spatial fluctuations in the background.  Recent models have begun to address the variations expected during hydrogen reionization \citep{mesinger08-ib, furl09-mfp}, demonstrating that they are crucial for interpreting measurements of that era and for understanding radiative feedback processes.

These fluctuations will persist even in the post-reionization era because of source clustering, stochastic fluctuations in the galaxy density, and radiative transfer effects.  Such variations are relatively small for the hydrogen-ionizing background (typically only a few percent of the mean value at $z \sim 3$; \citealt{zuo92a, zuo92b, fardal93, meiksin03, croft04}), principally because the mean free path of hydrogen-ionizing photons is extremely large ($\ga 100 \Mpc$; \citealt{madau99-qso, faucher08-ionbkgd}) and because sources (galaxies) are relatively common.  Fluctuations are larger at $z \ga 4$, when the mean free path is relatively small and galaxies are more highly clustered (e.g., \citealt{meiksin03}).  These studies concluded that such fluctuations are only significant if sources are rare and only measurable with future high-precision observations.  Thus the usual assumption of a spatially constant background is reasonable.

On the other hand, fluctuations in the helium-ionizing background have received less theoretical attention (though see \citealt{fardal98, maselli05, bolton06, meiksin07}).  Nevertheless, they are both larger and more observationally relevant than those for hydrogen.  One reason is that the IGM absorbs helium-ionizing photons more strongly than hydrogen-ionizing photons, leading to shorter attenuation lengths and larger fluctuations.  Second, \ion{He}{2} has an ionization potential of 54.4 eV, a sufficiently high energy that quasars (with hard spectra) are required to ionize it.  As such, the sources are quite rare, and themselves quite variable, implying large random fluctuations in the background.

Moreover, helium remains singly-ionized until $z \sim 3$, a regime that is relatively easy to observe.  A wealth of data now suggests dramatic evolution in the properties of intergalactic helium at about this time.  The strongest evidence comes from far-ultraviolet spectra of the \ion{He}{2} \lya forest along the lines of sight to several bright quasars at $z \sim 3$ \citep{jakobsen94, davidsen96, anderson99, heap00, smette02, zheng04, shull04, reimers04, reimers05, fechner06, fechner07}, which show a rapid increase in \lya absorption by \ion{He}{2} around that time -- although with substantial opacity fluctuations observed along several lines of sight \citep{anderson99, heap00, smette02, reimers05}.  These are most likely a direct effect of inhomogeneities in the helium-ionizing background during (or just after) reionization.  

Indirect measurements of the metagalactic ionizing background, which hardens as helium is reionized and the IGM becomes transparent to high-energy photons, suggests similar fluctuations after reionization.  In particular, measurements of the \ion{He}{2}/\ion{H}{1} ratio suggest that the helium-ionizing background fluctuates strongly at $z \sim 2.6$, with nearly order-of-magnitude spatial variations on scales spanning a few to a few tens of Mpc \citep{shull04, zheng04, fechner06, fechner07}.  Such strong fluctuations must be due to some combination of source count variations and radiative transfer effects.   \citet{meiksin07} showed with an analytic model that the broad distribution of quasar luminosities, together with their sparseness, accounts for much of the observed variation.  \citet{bolton06} also examined this regime with numerical simulations of quasar ionizing radiation, finding that the former effect could account for much, but not all, of the observed variance.  However, these simulations were limited to a relatively small box (less than a full attenuation volume).  \citet{maselli05} found (again using numerical simulations) that radiative transfer effects can also cause substantial variations, although they were hampered by the assumption of spatially uniform ionizing sources.  \citet{shull04} identified another contributing factor:  the broad distribution of quasar spectral indices, which directly affects the ionization rate.

Other observations of the \ion{He}{2}/\ion{H}{1} ratio suggest rapid time evolution as well \citep{heap00}, which is qualitatively consistent with helium reionization occurring at $z \sim 3$:  we would expect that the patchiness of the reionization process would lead to especially strong fluctuations between the bubbles of fully-ionized helium and the remaining \ion{He}{2} lying between them \citep{furl08-helium, mcquinn09}.  However, because of the difficulty of properly simulating helium reionization, these observations have not yet been quantitatively evaluated in light of modern reionization models.  The only previous study, by \citet{tittley07}, showed that the complex radiative transfer during \ion{He}{2} reionization could create wide variations in the hardness ratio.
 
These evolving variations should also be visible in metal lines.  For example, the ionization potentials of \ion{Si}{4} and \ion{C}{4} straddle that of \ion{He}{2}, so their ratio should evolve during helium reionization.  \citet{songaila98, songaila05} found a sudden break in that ratio at $z \sim 3$ (see also \citealt{boksenberg03}); modeling of the ionizing background from optically thin and optically thick metal line systems also shows a significant hardening at $z \sim 3$ \citep{vladilo03, agafonova05, agafonova07}.  However, other data of comparable quality show no evidence for rapid evolution \citep{kim02, aguirre04}.  These contrasting conclusions again suggest that the ionizing background may itself be fluctuating strongly between different lines of sight.

Here we study spatial and temporal variations in the helium-ionizing background using a combination of analytic and simple Monte Carlo models.  Moreover, we examine both the relatively simple post-reionization limit (as in \citealt{meiksin04}) and the behavior during reionization, when the fluctuations may be much stronger.  Extending the models to the reionization epoch allows us to predict the distribution of the hardness ratio, and hence the observability of the \ion{He}{2} \lya forest, during that era.  We will show that, in contrast to conventional wisdom, substantial transmission will remain throughout the bulk of helium reionization.

We describe our methods to compute the distribution of the amplitude of the ionizing background in \S \ref{method} and \ref{attenuation}.  We describe our post-reionization results in \S \ref{postreion} and those during reionization in \S \ref{during-reion}.  Finally, we conclude in \S \ref{disc}.

In our numerical calculations, we assume a cosmology with $\Omega_m=0.26$, $\Omega_\Lambda=0.74$, $\Omega_b=0.044$, $H_0=100 h \hunits$ (with $h=0.74$), $n=0.95$, and $\sigma_8=0.8$, consistent with the most recent measurements \citep{dunkley08,komatsu08}.   Unless otherwise specified, we use comoving units for all distances.

\section{Method}
\label{method}

We wish to compute the probability distribution $f(J)$ of the angle-averaged specific intensity of the radiation background at a given frequency (typically the ionization edge), $J$.  Its overall distribution in the IGM depends on four basic parameters.  The first two are the number density and luminosity distribution of ionizing sources (quasars in this case), parameterized by their comoving luminosity function $\Phi(L,z)$.  The third is only relevant during reionization:  the size of the local \ion{He}{3} bubble, $R$.  If $R$ is finite, the IGM surrounding the bubble will absorb ionizing photons from sources outside the local region.  The last parameter is the attenuation length for ionizing photons \emph{within} each \ion{He}{3} region, $r_0$.  We shall discuss $r_0$ in more detail in \S \ref{attenuation}; for now we treat it as a constant.

\subsection{The Quasar Luminosity Function}
\label{qsolum}

We use the recent estimate of the $B$-band quasar luminosity function $\Phi(L,z)$ over a broad range of redshifts from \citet{hopkins07}, who convert to a $B$-band luminosity function using the observed column density distribution of quasars selected in the X-ray (thus accounting for obscured sources).  This luminosity function is consistent with earlier estimates, but it does have a significantly flatter faint-end slope at higher redshifts, driven by recent measurements of faint $z \sim 3$ quasars \citep{hunt04, cristiani04, fontanot07, bongiorno07, siana08}.

At a given redshift, we define the dimensionless luminosity function $\phi(x) = \Phi(x L_\star,z)/n_i$, where $n_i$ is the total number density of ionizing sources,
\bq
n_i = \int_{L_{\rm min}}^{L_{\rm max}} dL \, \Phi(L,z),
\label{eq:phidefn}
\eq
$x=L/L_\star$, $L_\star$ is the mean luminosity of the sample,\footnote{Note that this is \emph{not} necessarily the same as the $L_\star$ parameter often used in fitting the quasar luminosity function.} and the integration ranges from minimum to maximum imposed luminosity cuts, $L_{\rm min}$ and $L_{\rm max}$.  The bright end of the luminosity function is sufficiently steep that our results are insensitive to the latter; the former is important because a large population of faint quasars provides a floor below which $J$ cannot fall.  We will typically use $L_{\rm min}=10^{43} \erg \, {\rm s}^{-1}$, slightly below the observational limit at $z \sim 3$.

We must next convert from these $B$-band luminosity functions to the photon energies of interest to us ($E>54.4 \eV$):  this requires a template for the spectral energy distribution of quasars.  We use
\bq
L_\nu \propto \left\{ 
\begin{tabular}{ll}
$\nu^{-0.3}$ & \qquad 2500 \mbox{ \AA} $< \lambda <$ 4400 \mbox{ \AA} \\
$\nu^{-0.8}$ & \qquad 1050 \mbox{ \AA} $< \lambda <$ 2500 \mbox{ \AA} \\
$\nu^{-\alpha}$ & \qquad $\lambda <$ 1050 \mbox{ \AA}.
\end{tabular}
\right.
\label{eq:qsotemplate}
\eq
At $\lambda > 1050$ \AA, this template agrees with that of \citet{madau99-qso}; other templates (e.g., \citealt{schirber03}) disagree in detail but do not affect our conclusions, given the uncertainties.  Most important for us is the far-ultraviolet spectral index $\alpha$.  At low redshifts, \citet{telfer02} find a wide variety of quasar spectral indices in the extreme ultraviolet, with a mean value of $\VEV{\alpha} \approx 1.6$.  This is slightly harder than the estimate of \citet{zheng98}, who found $\VEV{\alpha} \approx 1.8$, but shallower than the $\VEV{\alpha} \approx 0.5$ estimate of \citet{scott04} from $z<1$ quasars.  All these studies suggest substantial intrinsic source-to-source variance in $\alpha$, which affects the luminosity function above the helium ionization edge.  To model this, we assume that the probability distribution of $\alpha$ is Gaussian in the range $\alpha  \in (0.5,3.5)$, with mean $\bar{\alpha}=1.5$, and zero elsewhere; this provides a rough description of the \citet{telfer02} data.  We then convolve this distribution with the \citet{hopkins07} luminosity function for our calculations.  The variance in $\alpha$ slightly moderates the break near $L_\star$ and makes the bright end slightly shallower (from moderate luminosity sources with $\alpha < 1.5$), but it does not have a large effect on our results.

Note that our model for $f(J)$ therefore includes self-consistently the variations in $\alpha$ among the quasar population.  However, when we compare to observations we must further transform $J$ to the ionization rate $\Gamma$, which requires an integral over frequency and so has additional dependence on $\alpha$:  $\Gamma \propto (\alpha+3)^{-1}$ in the optically thin limit (see \S \ref{hard}; \citealt{shull04}).  We ignore this additional source of order-unity fluctuations, because our distributions are substantially broader anyway.

\subsection{Pure Attenuation:  The Analytic Model}
\label{analytic}

We now compute $f(J)$.  We first assume that helium reionization has completed; this corresponds to the limit $R \rightarrow \infty$.  In that case, the attenuation length $r_0$ fully describes the absorption of ionizing photons.  We further assume that Poisson fluctuations dominate variations in the quasar density; in that case, we can write the distribution analytically \citep{meiksin03}
\bqa
f_{R=\infty}(j) & = &  {1 \over \pi} \int_0^{\infty} ds \, \exp \left[ -s \bar{N}_0 \int dx \, x \phi(x) \, {\rm Im} \, G(sx) \right] 
\nonumber \\
& & \times \cos \left[ -sj + s \bar{N}_0 \int dx \, x \phi(x) {\rm Re} \, G(sx) \right],
\label{eq:jdist_r0}
\eqa
where $\bar{N}_0 = (4 \pi/3) n_i r_0^3$ is the mean number of sources in an attenuation volume, $j=J/J_\star$, $J_\star = L_\star/(4 \pi r_0)^2$,
\bq
G(t) = \int_0^\infty du \, \tau^3(u) e^{itu},
\label{eq:Gdefn}
\eq
and $u=e^{-\tau}/\tau^2$.  In this limit, the mean background is $\VEV{j}_{R=\infty} = 3 \bar{N}_0$, or $\VEV{J} \propto \bar{N}_0 J_\star \propto n_i L_\star r_0$.

\begin{figure}
\plotone{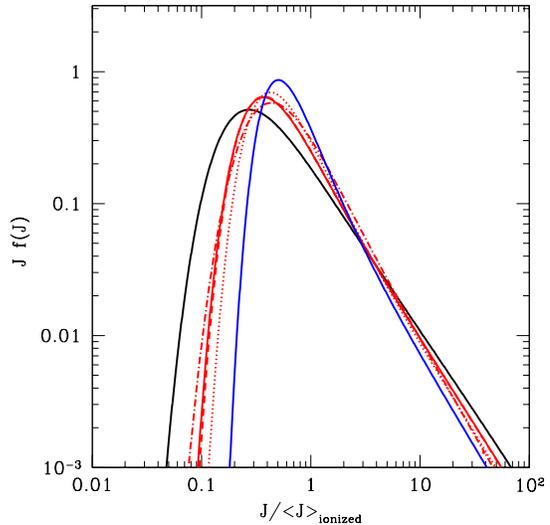}
\caption{Distribution of $J$ relative to its mean value in a fully-ionized IGM, $\VEV{J}_{\rm ionized}$.  The three solid curves assume $z=3$ and $r_0=25,\,35,$ and $55 \Mpc$, from widest to narrowest.  The other curves take $r_0=35 \Mpc$ but vary our other assumptions:  $L_{\rm min} = 10^{40} \erg \, {\rm s}^{-1}$ (dashed curve), fixing $\alpha=1.5$ (dotted curve), and assuming that all sources are $L_\star$ quasars with number density equal to $n_q(>L_\star)$ (dot-dashed curve).}
\label{fig:mw_ex}
\end{figure}

Figure~\ref{fig:mw_ex} shows some example distributions at $z=3$.  The three solid curves take $r_0=25,\,35,$ and $55 \Mpc$,\footnote{These values are comparable to the mean distance that a photon at the \ion{He}{2} ionization edge can travel from a quasar, and they also comparable to the mean spacing of \ion{He}{2} Lyman-limit systems at $z \sim 3$; see \S \ref{attenuation} for more details.}  from widest to narrowest (or from left to right in peak position).  The others take $r_0=35 \Mpc$ but vary some of our other assumptions.  The dotted curve assumes that all quasars have $\alpha=1.5$; it is slightly narrower than the fiducial model because of the reduced variance in the far-UV luminosity.  The dashed curve assumes that very faint quasars exist, with $L_{\rm min}=10^{40} \erg \, {\rm s}^{-1}$, three orders of magnitude fainter than the fiducial model.  This increases the source density by a factor of six but has almost no effect on $f(J)$, because the additional quasars contribute only a small fraction of the total emissivity.  Finally, the dot-dashed curve assumes that all quasars have $L=L_\star$ and normalizes their number density to $n = n_q(>L_\star)$ (so that, on average, only 0.6 quasars sit inside each attenuation volume).  In all cases $f(J)$ is very similar to our fiducial model.  Evidently, the shape of the distribution is fixed primarily by the number density of bright quasars.

Note in particular that the high-$J$ tail is nearly invariant.  These points correspond to regions near bright quasars (their ``proximity zones") and so depend only on the properties of that single bright source.  At sufficiently high $J$, $f(J) \propto J^{-5/2}$ \citep{meiksin03}.

Aside from the different quasar luminosity function and attenuation lengths, this calculation is identical to that in \citet{meiksin03} and \citet{meiksin07}, who also examined fluctuations in the post-reionization background.  Our results are similar to theirs, except that we find the low-$J$ tail of $f(J)$ to approach zero instead of asymptote to a finite value.  This has no practical importance for the observable results.

\subsection{Finite Bubbles:  Monte Carlo Approach}
\label{mc}

During reionization, we must include the additional parameter $R$,  the local \ion{He}{3} bubble's radius (or the horizon within which sources are visible).  \citet{meiksin03} describe how to obtain the full distribution $f(J)$ for arbitrary $R$ and $r_0$ (building upon the solution without attenuation first provided by \citealt{zuo92a}; see also \citealt{fardal93}).  However, we have found that the necessary integrals do not converge well when $\bar{N}$, the expected number of sources per bubble, is small.  Quasars are sufficiently rare that this is indeed the relevant limit:  for example, our fiducial model at $z=3$ has $\bar{N} \approx 1$ when $R \sim 22 \Mpc$ -- and, for $L_\star$ quasars, the expected number is unity only when $R \sim 40 \Mpc$.  (This contrasts sharply with the hydrogen reionization case, where $\bar{N}$ is always extremely large; \citealt{furl09-mfp}.)  Fortunately, it is precisely this rare source regime that is most amenable to a Monte Carlo treatment.

We first assume, as in \citet{meiksin03}, that the number of quasars is Poisson distributed around its mean value $\bar{N}$ in a region (see below for a discussion of deterministic clustering).  For a given $R$, we then randomly choose the number $N$ of quasars in the region.  Next, for the $i$th such quasar, we randomly assign a radius $r_i$ from the central point (assuming a uniform source density within the ionized region) and a normalized luminosity $x_i$.  (For simplicity, we always assume that the point of interest is at the center of the bubble, but we allow the sources to sit anywhere.)  Thus the specific intensity from all these quasars is
\bq
j = \sum_{i=1}^N {x_i \over (r_i/r_0)^2} e^{-r_i/r_0}
\label{eq:jcalc-mc}
\eq
where again $j = J/J_\star$.  The mean ionizing background is $\VEV{j} = 3 \bar{N}_0 (1 - e^{-R/r_0})$ \citep{meiksin03}.

There is one ambiguity in this approach:  the appropriate quasar number density to use when calculating $\bar{N}$.  The difficulty is that quasars \emph{must} sit inside of \ion{He}{3} bubbles.  A self-consistent reionization model would generate the bubbles around the sources, so this requirement would be manifestly satisfied.  However, in our Monte Carlo model we generate a bubble of a prescribed size and then randomly assign resident quasars.  If we simply used the \emph{observed} quasar number density $n_i$ to assign them to the bubbles, we would find a \emph{total} number density of $\bar{x}_{\rm HeIII} n_i$:  smaller than the observed value.

We therefore must renormalize the input source density to $\Phi(L)/\bar{x}_{\rm HeIII}$ in order to preserve the true \emph{total} quasar abundance and emissivity.  This effectively demands that quasars cluster in such a way that they are (uniformly) overabundant in ionized regions and absent in the rest of the IGM.  This does not accurately describe real quasar clustering, which is scale-dependent and sensitive to the details of the quasar host population, but it suffices for our purposes (especially given that stochastic fluctuations appear to dominate the topology of the quasar bubbles according to recent numerical simulations; \citealt{mcquinn09}).\footnote{This ambiguity does not occur for hydrogen reionization, where we can self-consistently compute the (clustered) halo population inside each ionized bubble \citep{furl04-lya,furl09-mfp}.}  However, when comparing distributions at different $\bar{x}_{\rm HeII}$, the effective source density (and average clustering) does differ (even if the comparison is done at fixed redshift).

We then generate $f(j)$ for a given set of $(R,r_0,z,\bar{x}_{\rm HeII})$ over $10^6$ trials.  Because each such trial is time-consuming, we construct a grid of distributions over $R$, with the other parameters fixed.  The grid spacing is $\Delta R=0.1 \Mpc$ at $R<20 \Mpc$, $\Delta R=0.5 \Mpc$ up to $R=35 \Mpc$, and $\Delta R=1 \Mpc$ to $R=70 \Mpc$.  Beyond that, we find it sufficient to use $\Delta R=10 \Mpc$ until we can approximate $f(j)$ with $f_{R=\infty}(j)$.  For intermediate values of $R$, we select the closest bubble size in our grid; we have verified that the spacing is small enough that interpolation is not necessary.

\begin{figure}
\plotone{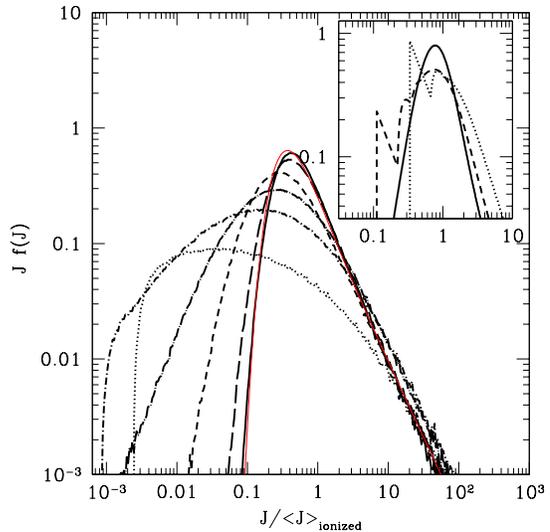}
\caption{Distribution of $J$ relative to its mean value in a fully-ionized IGM, $\VEV{J}_{\rm ionized}$.  All curves assume $z=3$, $r_0=35 \Mpc$, $\bar{x}_{\rm HeIII}=0.5$, and $L_{\rm min}=10^{43} \erg \, {\rm s}^{-1}$. The thick curves take $R=15,\,25,\,35,\,50,\,70,$ and $90 \Mpc$ from left to right in their peak locations (or dotted, short-dash-dotted, long-dash-dotted, short-dashed, long-dashed, and solid, respectively). The thin solid curve shows the distribution in a fully-ionized IGM.  \emph{Inset:}  Distributions for discrete bubbles with no attenuation, identical sources, and $\bar{N}=1,\,3$, and 10 (dotted, dashed, and solid curves, respectively.)}
\label{fig:mc_ex}
\end{figure}

Figure~\ref{fig:mc_ex} shows some example distributions for $z=3$, $r_0=35 \Mpc$, and $L_{\rm min}=10^{43} \erg \, {\rm s}^{-1}$.  The thick curves take $R=15,\,25,\,35,\,50,\,70,$ and $90 \Mpc$ from left to right in their peak location. Here, for concreteness, we take $\bar{x}_{\rm HeIII}=0.5$.  The thin solid curve shows the distribution in a fully-ionized IGM (from \S \ref{analytic}).    

Note that $\int f(J|R) dJ \neq 1$ in some of these cases.  This is because there is nothing in our formalism to demand that bubbles have at least one source; if they are empty, then $J=0$.  In practice, this is important for smaller bubbles (particularly $R \la 25 \Mpc$).  Physically, it would correspond to a case where a quasar appeared, ionized the region, and then faded away -- leaving a (mostly) ionized region with no active sources.  We will revisit such regions in \S \ref{fossil}.

As one might expect, the distributions generally become narrower as $R$ increases, because Poisson fluctuations become less important.  However, note that the smallest fluxes are not in the smallest bubbles:  the minimum non-zero flux in a discrete bubble is $L_{\rm min} e^{-R/r_0}/(4 \pi R)^2$, which actually decreases with $R$.  The Monte Carlo distributions clearly converge nicely to the \citet{meiksin03} limit for large bubbles:  this tail is filled by regions very close to a single quasar, so it should remain invariant as the bubble size changes.\footnote{Note that, for moderately sized bubbles, the Monte Carlo procedure overestimates the tail's amplitude; this is likely because of our simplified prescription that calculates the ionizing background at the center of each bubble; in reality, quasars can be up to two bubble radii away from the point of interest.  This has no effect on our final results.}  We note that $r_0$ has relatively little effect on $f(J)$ when $R$ is finite:  it is unimportant in small bubbles (because $e^{-R/r_0} \approx 1$), and large bubbles simply converge to the $R \rightarrow \infty$ distribution (which does not change dramatically with $r_0$; see Fig.~\ref{fig:mw_ex}).

\begin{figure}
\plotone{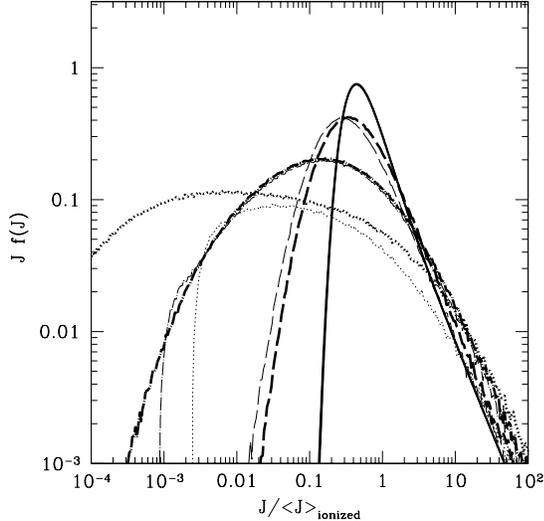}
\caption{Distribution of $J$ relative to its mean value in a fully-ionized IGM, $\VEV{J}_{\rm ionized}$.  All curves assume $z=3$, $\bar{x}_{\rm HeIII}=0.5$, and $r_0=35 \Mpc$. The dotted, dot-dashed, and dashed curves take $R=15,\,25,$ and $50 \Mpc$; the solid curves show the $R \rightarrow \infty$ limit.  The thin and thick curves assume $L_{\rm min}=10^{43}$ and $10^{40} \erg \, {\rm s}^{-1}$, respectively.  Note that the two solid curves overlap.}
\label{fig:mc_lumcomp}
\end{figure}

In fact, in contrast to the post-reionization limit, the parameter that most strongly affects the distribution is $L_{\rm min}$.  Figure~\ref{fig:mc_lumcomp} compares the distributions at $R=15,\,25,$ and $50 \Mpc$ (as well as the $R \rightarrow \infty$ limit) for $L_{\rm min}=10^{43}$ and $10^{40} \erg \, {\rm s}^{-1}$ (thin and thick curves, respectively).  Here we have again taken $\bar{x}_{\rm HeIII}=0.5$.  

As in the $R \rightarrow \infty$ limit, the extra sources make no difference for large bubbles, because they contribute only a small additional luminosity.  However, when $L_{\rm min}=10^{40} \erg \, {\rm s}^{-1}$ even small bubbles are likely to contain a source -- albeit with a low luminosity -- so $f(J)$ stretches to significantly smaller values.  It also has a larger total normalization.  Fortunately, this uncertainty has little real impact on our results, for two reasons.  First, as we will see below the integrated hard photon background from quasars outside the bubble will provide a minimum, nearly uniform ionizing background of $\sim 0.01$--$0.05 \VEV{J}$.  Second, we are most interested in the distribution fairly late in reionization, when most bubbles are large.  In that case, the variation with $L_{\rm min}$ is modest.  

As an aside, the broad luminosity function is also crucial for another reason.  The inset in Figure~\ref{fig:mc_ex} shows $f(J)$ in discrete bubbles with no attenuation if $\bar{N}=1,\,3,$ and 10 (dotted, dashed, and solid curves, respectively) and if all the sources are identical.  In this restricted case, \citet{zuo92a} has given an a closed-form solution for the distribution.  When $\bar{N} \sim 1$, $f(J)$ has discontinuities wherever an additional source can be added; for example, if $J < 2 L_{\rm min}/(4 \pi R)^2$, only one source can fill the bubble.  These discontinuities do not appear in our Monte Carlo distributions, because the wide allowed range of luminosities smoothes over the features.  They also disappear once $\bar{N} \ga 10$, and when $R \rightarrow \infty$, even if all sources have the same luminosity, because additional sources at arbitrary distances can fill in the gaps (see the dot-dashed curve in Fig.~\ref{fig:mw_ex}).

\subsection{Quasar Clustering}
\label{cluster}

The models described above assume that quasars are randomly distributed throughout the Universe; of course, in reality they sit inside massive dark matter halos that cluster strongly.  We can gauge the importance of the variance induced by this deterministic clustering by computing the typical standard deviation in the number of quasar halos per attenuation volume, $\sim \bar{b} \sigma(r_0)$, where $\bar{b}$ is the average bias of quasar host halos and $\sigma^2(r_0)$ is the variance in the dark matter field smoothed over a radius $r_0$ at the appropriate redshift.  At $z=3$, $\sigma \approx (0.19,0.11)$ for $r_0=(25,45) \Mpc$ in our cosmology.  Recent $z \sim 3$ surveys have found $\bar{b} \sim 8$ for bright quasars in the SDSS survey \citep{shen07} and $\bar{b} \sim 5.5$ for intermediate-luminosity quasars selected via X-rays \citep{francke08}.  We will use the latter as more representative of the luminosity range over which most of the emission is produced.  In that case, $\bar{b} \sigma(r_0) \sim 0.6$--1, and we expect clustering to induce order-unity fluctuations in the ionizing background.  This probably underestimates the importance of clustering, because clustering \emph{within} the attenuation volumes cannot really be neglected.  This may be especially important in the high-$J$ tail, where an over-abundance of close quasar pairs may increase the amplitude of $f(J)$.

Fortunately, even a cursory glance at Figure~\ref{fig:mc_ex} suffices to show that -- even after reionization is complete -- the Poisson fluctuations in the number of visible quasars, together with their wide luminosity distribution, provide much larger variations.  We have verified that this is a reasonable approximation by convolving our fiducial post-reionization distribution with the underlying halo density distribution (which fixes $\bar{N}_0$ in each volume element).  This behavior differs strongly from hydrogen reionization, where the Poisson fluctuations are tiny and clustering is essential to estimating the fluctuations.

\section{The Attenuation Length}
\label{attenuation}

One of the crucial inputs to our formalism is the attenuation length of ionizing photons, $r_0$.  Here we will describe how we estimate this length scale; we will follow the method originally presented in \citet{furl08-helium} and refer the reader there for more details.

To estimate the attenuation length around an individual quasar, we need to find the distance at which a system with $\tau \sim 1$ typically lies.  We therefore need to associate overdensities with column densities in order to track the total amount of absorption within each system.  \citet{schaye01} has shown that \ion{H}{1} \lya forest absorbers can be accurately modeled by assuming their physical scale to be comparable to the Jeans length $L_J \approx c_s/(G \rho)^{1/2}$.  We will use the same approximation for helium, so that an absorber with overdensity $\Delta = \rho/\bar{\rho}$ has column density $N_{\rm HeII} \approx L_J x_{\rm HeII} n_{\rm He}$.  Then
\bq
N_{\rm HeII} \approx 1.8 \times 10^{15} \Delta^{3/2} T_4^{-0.2} \left( {1+z \over 4} \right)^{9/2} \left( {10^{-14} \secinv \over \Gamma} \right),
\label{eq:nheii}
\eq
where we have assumed photoionization equilibrium, used the case-A recombination rate, and $T = 10^4 T_4 \kel$ is the IGM temperature.  A system will become optically thick at a frequency $\nu$ when $\tau_\nu = \sigma_\nu N_{\rm HeII} = 1$.  At a distance $R$ from a quasar with $B$-band luminosity $L_B$, this requires an overdensity $\Delta_i$
\bq
\Delta_i \approx 64 T_4^{2/15} \left( {L_B \over 10^{12} \Lsun} \right)^{2/3} \left( {{\rm Mpc} \over R} \right)^{4/3} \left( {1+z \over 4} \right)^{-5/3}.  
\label{eq:deltai}
\eq
at the ionization edge.  Higher-energy photons can penetrate even denser systems than implied by this simple model (see \S \ref{hard} below), but the photoionization cross section $\sigma_\nu = \sigma_{\nu_{\rm HeII}} (\nu_{\rm HeII}/\nu)^3$ (with $\sigma_{\nu_{\rm HeII}} = 1.91 \times 10^{-18} \cmsq$), so the instantaneous ionization rate (the most important quantity for us) is dominated by low-energy photons.  However, effects that accumulate over time -- such as heating -- depend more on the hard photons \citep{abel99, bolton08, mcquinn09}.

Systems with $\Delta > \Delta_i$ are the ``Lyman-limit systems" that determine $r_0$.  In the limit of a uniform ionizing background, $r_0$ will be the mean separation of such overdensities; this is the relevant calculation during hydrogen reionization, for example \citep{furl05-rec, choudhury08, furl09-mfp}.  However, we have already seen that the rarity of bright quasars induces large fluctuations in the helium-ionizing background, especially when the ionized bubbles are still separated by neutral walls.  In this case, most points in the IGM are illuminated by only one (or at most a few) quasars.  Then it is better to compute the maximum distance from each such quasar that an ionizing photon can travel.  

Thus, to obtain $r_0$, we simply find the point at which the  mean separation of the neutral blobs, $\lambda_i(\Delta_i)$, equals the distance from the quasar (note that $\lambda_i$ decreases with distance from the quasar because its ionizing intensity decreases); this is the distance at which a photon is likely to have encountered a system neutral enough to absorb it.  To compute the mean separation between IGM patches with $\Delta = \Delta_i$, we employ the IGM density distribution from \citet{miralda00}, as well as their prescription for $\lambda_i(\Delta_i)$, which provides a good fit to numerical simulations at $z \sim 2$--$4$.  More recent simulations show deviations from this fit at high densities, so our results should only be viewed as a rough guide \citep{pawlik09,bolton09-density}; for this reason, we examine a relatively wide range in attenuation length.  Moreover, we ignore a few other important effects, like the accumulated photoelectric absorption of the low-density IGM; see below for a discussion of these problems.

\begin{figure}
\plotone{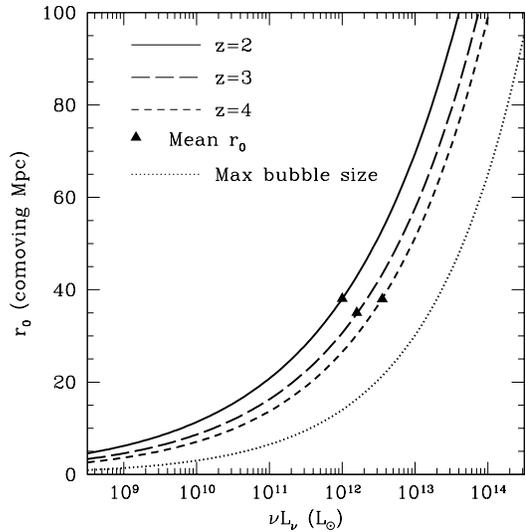}
\caption{Attenuation length as a function of $B$-band quasar luminosity, assuming that each quasar is isolated.  The solid, long-dashed, and short-dashed curves are for $z=2, \, 3,$ and $4$, respectively.  The filled triangles mark the luminosity-weighted mean $\VEV{r_0}$ for each redshift.  The dotted curve shows the maximum radius of a \ion{He}{3} bubble around an isolated quasar, neglecting recombinations and assuming a lifetime of $10^7 \yr$.}
\label{fig:rmax}
\end{figure}

Figure~\ref{fig:rmax} shows $r_0(L)$ for quasars at $z=2,\,3,$ and $4$ (solid, long-dashed, and short-dashed curves, respectively).  The filled triangles indicate the luminosity-weighted mean $\VEV{r_0}$ across the entire quasar population.  We find that  $\VEV{r_0} \approx 35$--$38 \Mpc$ over this redshift range; the increasing clumpiness, increasing mean luminosity, and decreasing mean IGM density roughly cancel each other out.  For reference, the dotted curve shows the maximum size of the ionized bubble surrounding an isolated source (assuming a quasar with a lifetime of $10^7 \yr$ shining into \ion{He}{2}), computed by equating the total number of ionizing photons to the total number of \ion{He}{2} ions in the bubble (thus ignoring recombinations).  Even if quasars live for $10^8 \yr$, $r_0$ will still be comparable to the radius of each quasar's ionization zone.  Thus, we expect attenuation to be relatively unimportant early in reionization but to become more significant as quasars are born into larger pre-ionized regions.  

Once the ionizing background becomes more uniform, we must include the accumulated background from more distant sources when estimating where neutral absorbers can appear.  By relating the helium absorbers to the hydrogen \lya forest, one can show that a uniform background yields $\lambda_{\rm He} \sim 6.6,\,12,$ and $30 \Mpc$ at $z=4,\,3,$ and 2, with at least a factor of two uncertainty from the amplitude of the helium-ionizing background \citep{furl08-helium}.  This suggests that the transition from single-source to uniform is probably gradual and relatively smooth.  Our estimates also compare well to others in this regime:   for example, \citet{bolton06} take $r_0 = 30 [(1+z)/4]^{-3} \Mpc$.  

With the luminosity function and attenuation length in hand, we can now perform an important self-consistency check:  do these parameters reproduce the observed mean \ion{He}{2}-ionizing background, often measured through the hardness ratio (see \S \ref{postreion} below)?  With the \citet{hopkins07} luminosity function and $R_0 \sim 40 \Mpc$ at $z=2.5$, and with $\VEV{\alpha}=1.5$, we find $\Gamma \approx 6 \times 10^{-15} \secinv$.  This is comparable to the simulations of \citet{bolton06}, which found reasonable agreement to the observations when a fluctuating ionizing background is included.  It is certainly well within the errors of the measurements (including that of $\VEV{\alpha}$, which causes about a factor of two uncertainty in the total emissivity).  We consider this agreement adequate, but note as well that when comparing to the observations we typically normalize the mean background to the observations, which is equivalent to adjusting $\VEV{\alpha}$.

In detail, our procedure ignores a few important effects.  First, we overestimate $\lambda_i$ by up to a factor $\sim 2$ because of the accumulated photoelectric absorption of optically thin systems \citep{furl05-rec}.    Second, we have computed the mean free path at an arbitrary point in the IGM, whereas we are actually interested in the mean free path as seen by a quasar, which most likely sits in an overdense region (see also \citealt{yu05, alvarez07, lidz07}).  This will also cause us to overestimate the mean free path.  However, even at $z \sim 6$ (where these massive halos are much more rare) the environments typically approach the mean density within $\la 20 \Mpc$ of the quasar, so it should not be a large effect.  Finally, we have ignored higher energy photons, which can travel much farther.  We will now consider their accumulated background.

\subsection{The Hard Photon Background}
\label{hard}

To this point, we have assumed that, during reionization, all the incident ionizing radiation on a point comes from quasars within the local ionized volume, i.e. the \ion{He}{2} in between fully-ionized bubbles absorbs all the radiation from external sources.  However, high-energy photons can propagate large distances, even if all of the helium remains singly-ionized:  for a uniform IGM, the mean free path is
\bq
\lambda_u = \lambda_{\rm edge} \bar{x}_{\rm HeII}^{-1} \left( {\nu \over \nu_{\rm HeII}} \right)^3 \left( {1+z \over 4} \right)^{-2},
\label{eq:mfp-uniform}
\eq
where $\lambda_{\rm edge} = 0.72 \Mpc$ is the mean free path at the ionization edge.  High-energy photons will therefore create a diffuse, nearly uniform background that provides a lower limit to $J$, even before reionization is complete.

To estimate this limiting value, we first assume that all the ionized bubbles have a fixed size and a spatial number density $n_{\rm bub} \sim \bar{x}_{\rm HeIII}/(4 \pi R^3/3)$.  Assuming (naively) that these bubbles are randomly distributed, their average spacing will be $\Delta \ell \sim 1.6 R/\bar{x}_{\rm HeIII}^{1/3} \sim R$ during the middle phases of reionization.  For a simple estimate, we therefore assume that all photons with $\lambda_u < \lambda_u(\nu_{\rm thin}) \approx (\Delta \ell - R)$ are blocked by \ion{He}{2} regions in the IGM, while others traverse a distance $\lambda_u(\nu)$ before being blocked.  Thus $\nu_{\rm thin}$ is the minimum frequency for which photons are able to traverse one of the \ion{He}{2} regions between the fully-ionized bubbles.

We now wish to compare the ionization rate due to these high-energy photons, $\VEV{\Gamma}_{\rm thin}$, to the mean ionization rate from quasars within a discrete bubble (ignoring all sources outside of that bubble, and ignoring attenuation within the bubble for simplicity).  This latter quantity is
\bqa
\VEV{\Gamma}_{\rm bubble} & = & \VEV{J(\nu_{\rm HeII})}_R \int_{\nu_{\rm HeII}}^\infty {d \nu \over h \nu} \left( {\nu \over \nu_{\rm HeII}} \right)^{-\alpha} \sigma_\nu \\
& = & \frac{ \sigma_{\nu_{\rm HeII}} }{h (\alpha + 3)} \VEV{J(\nu_{\rm HeII})}_R ,
\label{eq:gamma-thick}
\eqa
where $\VEV{J_{\nu_{\rm HeII}}}_R = 3 \bar{N} L_\star(\nu_{\rm HeII})/(4 \pi R)^2$ is the mean specific intensity inside a bubble, evaluated at the helium ionizing edge $\nu_{\rm HeII}$ and ignoring attenuation \citep{zuo92a}, $\bar{N} = (4 \pi/3) R^3 n_i$ is the mean number of sources contained in a bubble, and where we have assumed that all quasars have the mean spectral index $\alpha$.

In contrast, the ionization rate from the uniform, high-energy background is 
\bqa
\VEV{\Gamma}_{\rm thin} & = & \VEV{J(\nu_{\rm thin})}_{R=\infty} \int_{\nu_{\rm thin}}^\infty {d \nu \over h \nu} \left( {\nu \over \nu_{\rm thin}} \right)^{-\alpha} \sigma_\nu  \nonumber \\ 
& & \times \left[ {\lambda_u(\nu) \over \lambda_u(\nu_{\rm thin}) } \right], \\
& = &  \frac{ \sigma_{\nu_{\rm HeII}} }{h \alpha} \VEV{J(\nu_{\rm thin})}_{R=\infty} \left( {\nu_{\rm HeII} \over \nu_{\rm thin}} \right)^{3}, 
\label{eq:gamma-thin}
\eqa
where $\VEV{J(\nu_{\rm thin})}_{R=\infty} = 3 \bar{N}_0 L_\star(\nu_{\rm thin})/[4 \pi \lambda_u(\nu_{\rm thin})]^2$ is the radiation background evaluated at the frequency where the IGM becomes optically thin.  The factor in square brackets accounts for the increased mean free path of high-energy photons (see eq.~\ref{eq:mfp-uniform}).\footnote{By ignoring IGM clumping, we conservatively overestimate the background from optically thin photons:  the mean free path may increase more slowly than eq.~(\ref{eq:mfp-uniform}) in the presence of dense, self-shielded systems that block high-energy photons relatively efficiently (as in the hydrogen \lya forest); see \citet{bolton08, mcquinn09}.}  Note in equation~(\ref{eq:gamma-thin}) the factor $({\nu_{\rm HeII} / \nu_{\rm thin}})^{3}$, which shows the suppression of the ionization rate at high frequencies.

Thus we find
\bqa
{\VEV{\Gamma}_{\rm thin} \over \VEV{\Gamma}_{\rm bubble}} & = & \left( {\alpha + 3 \over \alpha} \right) \left[ { \lambda_u(\nu_{\rm thin}) \over R } \right] \left( {\nu_{\rm HeII} \over \nu_{\rm thin}} \right)^{3+\alpha} 
\label{eq:gamma-ratio} \\
& \approx & \left( {\alpha + 3 \over \alpha} \right) \left( {\lambda_{\rm edge} \over R} \right)^{1+\alpha/3} \left( {R \over \lambda_{\rm thin}} \right)^{\alpha/3} \nonumber \\
& & \times \left( {1+z \over 4} \right)^{-2 -2 \alpha/3} \bar{x}_{\rm HeII}^{-1 - \alpha/3}.
\eqa
The factor with $\lambda_{\rm thin}$ can be re-expressed as $\approx (1.6 \bar{x}_{\rm HeIII}^{-1/3} - 1)^{-\alpha/3}$, which is close to unity.  We then find that the ratio is $\la 0.1$ for $\alpha=1.5$ and $R=15 \Mpc$ at $z=3$, so long as $\bar{x}_{\rm HeII} \la 0.6$; at larger ionized fractions, the bubbles become several times larger \citep{furl08-helium}, so the contribution remains at a few percent through most of reionization.  (The ratio in eq.~\ref{eq:gamma-ratio} becomes large very near the end of reionization, because this simple model ignores attenuation from dense clumps.)

Thus the integrated high-energy background only provides a few percent of the total ionization rate at any point during reionization.  This is primarily because $\sigma_\nu \propto \nu^{-3}$:  although the radiation intensity can be reasonably large, only a small fraction of the photons actually interact.  Again, effects that accumulate over long time intervals (such as heating) can still be quite significant \citep{abel99, bolton08, mcquinn09}.  

The high-energy photons are more important after reionization is over, because then $\nu_{\rm thin} \rightarrow \nu_{\rm HeII}$.  In that case, assuming Poisson-distributed absorbers, we can approximate the attenuation length as $r_0 \propto \nu^{\beta}$; $\beta=3/2$ for a column density distribution $\propto N^{-3/2}$,  a reasonable approximation for the \ion{H}{1} \lya forest \citep{paresce80, zuo93, lidz07}.  (The scaling may approach $\nu^{-3}$ as in eq.~\ref{eq:mfp-uniform} for photons well above the ionization ege, but again they contribute only a modest amount to the total ionization rate; see \citealt{bolton08, mcquinn09}.)  We could then estimate the net fluctuations by convolving $f(J|r_0)$ with the spectrum-weighted ionization cross-section; we will take a simpler approach and estimate $r_0$ at a characteristic ionization frequency.  The ionization rate from a logarithmic frequency interval around $\nu$ is $\nu \Gamma_\nu \propto \nu (L_\nu/h \nu) \sigma_\nu r_0(\nu) \propto \nu^{-(3 + \alpha - \beta)}$.  The effective frequency is therefore
\bq
\nu_{\rm eff} = \frac{\int_{\nu_{\rm HeII}}^\infty d\nu \, \nu \Gamma_\nu}{\int_{\nu_{\rm HeII}}^\infty d \nu \Gamma_\nu} = {(3 + \alpha - \beta) \over (2 + \alpha - \beta)} \nu_{\rm HeII},
\label{eq:nu-eff}
\eq
or $r_0(\nu_{\rm eff})/r_0(\nu_{\rm HeII}) = (\nu_{\rm eff}/\nu_{\rm HeII})^\beta = 1.84$ for $\beta=\alpha = 3/2$.  Note, however, that a full radiative transfer calculation provides a much more complex spectrum, especially when recombination radiation is included, so these power-law estimates will not be particularly accurate in some regimes \citep{haardt96, fardal98}.

This increased mean free path can make a factor $\sim 2$ difference to $f(J)$, as shown by the solid curves in Figure~\ref{fig:mw_ex}.  We note, however, that the variance in $f(J)$ is a nonlinear function of the attenuation length; in particular, fluctuations will be strongest at the ionization edge, which also contributes most strongly to $\Gamma$.  Thus this probably underestimates the real variation; for concreteness we will use $r_0=35 \Mpc$ as our fiducial value below, comparable to the values from Figure~\ref{fig:rmax} and to estimates based on the \lya forest.

\section{The Ionizing Background After Helium Reionization}
\label{postreion}

As a first application of this method, we consider variations in the ionizing background after helium reionization is complete.  In this regime, our analytic model from \S \ref{analytic} describes the distribution of $f(J)$ and hence $f(\Gamma)$ (once an appropriate average attenuation length is chosen).\footnote{There is one additional complication:  scatter in $\alpha$ directly affects the ionization rate through the high-energy photons (eq. \ref{eq:nu-eff}), over and above the additional luminosity dispersion that we included in \S \ref{qsolum} \citep{shull04}.  We do not include this additional scatter, although it only introduces extra fluctuations of order unity.} For example, Figure~\ref{fig:mw_ex} showed some post-reionization distributions at $z=3$.  Regardless of our assumptions about the minimum quasar luminosity and the attenuation length, $f(\Gamma)$ has a large variance.  We parameterize the variation by the range $\Delta \Gamma$ for which $f(\Gamma) > f(\VEV{\Gamma})/2$.  For the curves in Figure~\ref{fig:mw_ex} with $r_0=(25,\,35,\,55) \Mpc$, we find $\Delta \Gamma/\VEV{\Gamma} = (1.3,\, 1.15,\, 1.07)$.  Order unity fluctuations are typical, and a non-negligible fraction of points can have much larger ionizing backgrounds.  (Note, however, that the \emph{median} $\Gamma$ is smaller than the mean.)

This distribution should be observable via the \lya forest; in particular, by combining information about the hydrogen and helium forests we can measure the fractional fluctuations in the ionizing backgrounds for \ion{H}{1} and \ion{He}{2}.  The former is most strongly affected by photons near the hydrogen-ionization edge; at $z \sim 2$--$3$, their mean free path is hundreds of Mpc \citep{madau99-qso, faucher08-ionbkgd}, and each attenuation volume contains a huge number of quasars and star-forming galaxies.  We therefore assume that the hydrogen-ionizing background is truly uniform \citep{meiksin03, croft04}.  In that case, the \ion{He}{2}/\ion{H}{1} ratio serves as a proxy for $f(\Gamma)$ for helium.

The equivalent \citet{gunn65} IGM optical depth due to \ion{He}{2} in a parcel of gas with relative overdensity $\Delta = \rho/\bar{\rho}$ is 
\bq
\tau_{\rm HeII} = 3.55 \times 10^3 \, x_{\rm HeII} \Delta \left( {1+z \over 4} \right)^{3/2}.
\label{eq:taugp}
\eq
In ionization equilibrium with a helium-ionizing background of amplitude $\Gamma = 10^{-14} \Gamma_{14} \secinv$, this is
\bq
\tau_{\rm He II} = 8.0 \Delta^2 \Gamma_{14}^{-1} \left( {1+z \over 4} \right)^{9/2} \left[ {\alpha_{\rm HeII} \over \alpha_A(1.5 \times 10^4 \kel) } \right],
\label{eq:tauioneq}
\eq
where we have used case-A recombination under the assumption that most of the recombination photons will escape the local volume and ionize denser regions (e.g., \citealt{miralda03}).

Variations in $\tau_{\rm HeII}$ can therefore be traced to density structure in the IGM, temperature variations, or the ionizing background.  We will ignore temperature fluctuations for simplicity, though they may also be important if \lya forest lines are thermally broadened by high post-reionization temperatures \citep{gleser05, furl08-igmtemp, mcquinn09}.  Density variations can be calibrated by comparison to the \ion{H}{1} \lya forest, so we can isolate the effect of $\Gamma$ by measuring $\eta = N_{\rm HeII}/N_{\rm HI}$, which is (nearly) directly observable by comparing lines in the hydrogen and helium \lya forests.  Assuming the gas is highly-ionized and in photoionization equilibrium, we have
\bq
\eta = {n_{\rm HeIII} \over n_{\rm H II}} {\alpha_{\rm HeII} \over \alpha_{\rm HI}} {\Gamma_{\rm HI} \over \Gamma} \equiv {\eta_0 \VEV{\Gamma} \over \Gamma}.
\label{eq:eta}
\eq  

We normalize the distribution of $\eta$ by fixing its mean to a specified value.  Unfortunately, this choice is not straightforward.  Theoretical estimates span $\VEV{\eta} \sim 40$--80 \citep{haardt96, fardal98}.  \citet{bolton06} found that setting $\VEV{\eta} \approx 60$ reproduces the observed optical depth in the helium \lya forest when a fluctuating radiation field is included.  Observations show a wide scatter in the values of $\eta$ in individual systems, but the average is $\sim 45$--$110$ \citep{shull04,zheng04, fechner06, fechner07}, with most of the uncertainty due to systematics in interpreting the spectra and in how to treat outliers.  For concreteness, we will set $\VEV{\eta} = 80$, for easy comparison to the data of \citet{fechner06}.  Note, however, that this is somewhat larger than estimates from, e.g., \citet{shull04}.

\begin{figure}
\plotone{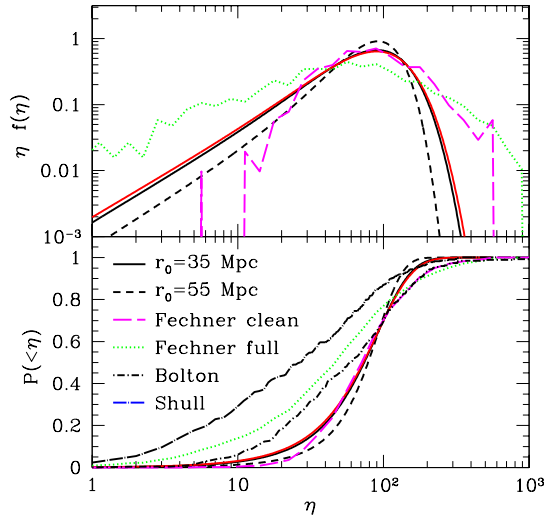}
\caption{Differential and cumulative probability distributions of the hardness ratio, $\eta$, assuming that helium reionization has ended (top and bottom panels, respectively).  The two solid curves take $z=3$ and $2.5$, assuming $r_0=35 \Mpc$.  The short-dashed curve assumes $r_0=55 \Mpc$, respectively.  The long-dashed and dotted curves show the ``clean" subsample and full data set of \citet{fechner06}, respectively.  In the bottom panel, the long dash-dotted curve shows the observational data from \citet{shull04}, and the short dash-dotted curve shows the simulated distribution of \citet{bolton06}.}
\label{fig:feta_post}
\end{figure}

Figure~\ref{fig:feta_post} shows the resulting differential and cumulative distributions of the hardness ratio in the post-reionization limit.  The solid curves assume $r_0=35 \Mpc$, at $z=3$ and 2.5.  The variations over this redshift range are extremely small, because the quasar luminosity function is not evolving rapidly ($\bar{N}_0 \sim 13$--$19$ from $z=3$--$2$, if $r_0$ is constant).  Thus any observed evolution in $f(\eta)$ over this interval would indicate evolution in the mean free path.  

The short-dashed curve shows $f(\eta)$ for $r_0=55 \Mpc$ at $z=2.5$.  Just like $f(\Gamma)$, the hardness parameter has a larger variance if the attenuation length is smaller, because Poisson fluctuations in the quasar counts become more important.

The short dash-dotted curve in the bottom panel shows the cumulative distribution from the \citet{bolton06} simulations, which included discrete, clustered sources within each attenuation volume but did not model radiative transfer through a clumpy IGM.  Their distribution is somewhat broader than ours (note that we have renormalized the values to our choice of $\VEV{\eta}$).  One reason is their treatment of attenuation: their simulation box was $\sim 30 h^{-1} \Mpc$ across, comparable to their assumed $r_0$.  To simplify the effects of absorption, they only allowed sources within one attenuation length to illuminate any given point, but they ignored absorption within that region.  This likely accounts for at least part of the discrepancy, because the added sources at large distances will help to damp out smaller scale fluctuations (for example, compare the dot-dashed and thin solid curves in Fig.~\ref{fig:mc_ex}).  The rest is likely due to the much more realistic treatment of \lya forest features in the simulations.

\subsection{Comparison to Observations} \label{obs}

The long-dashed and dotted curves show the observed $\eta$ distributions of \citet{fechner06}.  The dotted curve is their complete sample; the long-dashed curve is their more reliable sample of moderate optical depth lines, where the comparison of \ion{H}{1} and \ion{He}{2} columns can be made most reliably (these systems have $0.01 \le \tau_{\rm HI} \le 0.1$).  The agreement between our model and this restricted sample is quite impressive:  the only significant discrepancy is a longer tail toward high $\eta$ in the data.  This may arise partly from an underestimate of $\VEV{\eta}$; \citet{meiksin07}, who performed a similar comparison (and obtained results similar to ours), found that such a modification would improve agreement at that end.

On the other hand, the full data set shows relatively poor agreement, with significant tails toward both large and small $\eta$.  Other data show similar trends; for example, the long dash-dotted curve in the bottom panel shows the observed cumulative probability distribution from \citet{shull04}, using FUSE measurements of the \ion{He}{2} forest (smoothed on 0.2 \AA \, scales) and high-quality optical data of the \ion{H}{1} \lya forest along the line of sight to HE 2347--4342; the data span the range $z=2.3$--2.9.  They claim to measure $\eta$ in the range 0.1--460 reliably; they have censored points within $1 \sigma$ of either zero or complete transmission, which does complicate the comparison.\footnote{In particular, the mean of this censored data set is $\VEV{\eta}=45$, smaller than our distribution. Thus the censored points lie primarily at very large $\eta$, extending the observed distribution to the right of the plot.    The differences with our models described in the text are qualitatively unchanged if we rescale our models to match this mean.}  \citet{zheng04} independently analyzed the same data with different techniques (line-fitting, as opposed to pixel methods) and obtained a similar distribution.

Both of these full distributions are considerably broader than the theoretical predictions, for any reasonable attenuation length, and the \citet{fechner06} subsample.  This is not surprising, because a variety of effects other than shot noise in the quasar counts contribute to the observed distribution, including observational errors, clustering, temperature fluctuations, line broadening and peculiar velocities in the forest, extra fluctuations from variations in quasar spectral indices, and radiative transfer effects through the IGM \citep{abel99, bolton08}.  Moreover, especially with the \citet{shull04} data, the observations span a relatively large redshift interval, over which $\VEV{\eta}$ evolves significantly.  

Overall, the agreement with the ``clean" data seems remarkably good, as it was in the simpler model of \citet{meiksin07}.  Shadowing and similar radiative transfer effects may be responsible for the high-$\eta$ tail, which corresponds to a small \ion{He}{2} ionizing background \citep{tittley07}.  The good agreement with this restricted set suggests that the largest problem with the comparison to the full data sets arises from observational errors.

Two other aspects of the data merit some discussion.  First, \citet{shull04} and, with less confidence, \citet{fechner06} detected an anti-correlation between $N_{\rm HI}$ and $\eta$, so that void-like regions in the IGM tended to have softer ambient radiation fields.  \citet{fechner06} and \citet{fechner07} showed that this is at least partly due to noise in low-column density systems and saturation in higher-column density systems, and they also argued that the anti-correlation would disappear if higher-column density lines were thermally, rather than turbulently, broadened.  In any case, our model cannot address such a trend, as we do not include the density field in our calculations.  The best we can say is that dense absorbers are most likely to be near the massive quasar hosts, which would probably cause a weak association with low-$\eta$ lines.  Naively, voids should be preferentially illuminated by a softer, more highly-attenuated radiation field because they are farther from the sources, at least if the hydrogen-ionizing background itself is uniform \citep{bolton06}.  Radiative transfer, on the other hand, may cause a correlation in the opposite direction \citep{maselli05}.

Second, \citet{shull04} detected substantial $\eta$ fluctuations on rather small scales ($\sim 2 \Mpc$) toward HE 2347--4342.  The natural fluctuation scale in our model is the attenuation length, $r_0$.  However, \citet{fechner07} showed that the highest signal-to-noise portions of that spectrum, and the entire line of sight to HS 1700+6416 (also high signal-to-noise), show much smoother variations, with $\sim 33\%$ of the IGM varying on scales $\la 6 \Mpc$ and $\ga 50\%$ varying only on scales $\ga 14 \Mpc$.  The latter is not so far from our expected attenuation lengths, and additional effects such as radiative transfer, density structure in the IGM, and aliasing may explain the differences.  

\section{The Ionizing Background During Helium Reionization}
\label{during-reion}

\subsection{The Bubble Size Distribution}
\label{bubble-size}

We now turn to the helium ionizing background before overlap, when portions of the IGM are still filled by \ion{He}{2}.  In this case, our model requires one additional ingredient:  the size distribution $n_b(R)$ of discrete ionized bubbles at a given time during reionization.  

For this purpose, we use the excursion-set approach of \citet{furl08-helium}, which is based on a model originally developed for hydrogen reionization in \citet{furl04-bub}.  In brief, the model generates \ion{He}{3} regions by comparing the number of ionizing photons generated inside the region to the number of helium atoms; large-scale overdensities host more massive halos (and hence more quasars) and so ionize themselves earlier.  The model uses the excursion set formalism to compute the scales at which any given IGM parcel enters an ionized region, and this is interpreted as $n_b(R)$.

The two inputs for this model are $\bar{x}_{\rm HeIII}$ and a prescription for the helium-ionizing fluence of dark matter halos (essentially, a model for placing quasars in their host galaxies); the resulting $n_b(R|\bar{x}_{\rm HeIII})$ is nearly independent of redshift.  As a fiducial model, we will assume that halos with $m \ga 5 \times 10^{11} \Msun$ host quasars, and that the total fluence is proportional to the halo mass.\footnote{Note that this assumption does not affect the quasar number density, which we fix to the observed value.}  This provides a reasonable match to clustering measurements of luminous, high-$z$ quasars and to the measured redshift dependence of the total quasar emissivity \citep{furl08-helium}.  Physically, it assumes that every massive halo contains a supermassive black hole with a fixed radiative efficiency.

Although the excursion-set model, which is driven by deterministic halo clustering, works very well for hydrogen reionization (in comparison to numerical simulations; \citealt{zahn07-comp, mcquinn07, mesinger07}), it is not yet clear how successful it will be with helium reionization, due to several complications during the latter era.  First, quasars are much rarer than the galaxies that (probably) dominated hydrogen reionization, and the same Poisson fluctuations described in our model for $f(J)$ will also dominate the topology of ionized gas during the early (and perhaps middle) stages of reionization.  We mimic this by setting a minimum bubble size $R_{\rm min}$ during the early stages, when the rare quasars are scattered about nearly randomly \citep{furl08-helium}.  We set $R_{\rm min}=15 \Mpc$, appropriate for an $L_\star$ quasar shining for $10^7 \yr$ (see the dotted curve in Fig.~\ref{fig:rmax}).

Second, the relatively hard spectra of quasars allow some high-energy photons to propagate large distances through the IGM before being absorbed (as in \S \ref{hard}).  These will gradually ionize and heat regions far from quasars.  Thus it is not formally correct to divide the IGM into pure \ion{He}{2} and pure \ion{He}{3}, as the excursion-set model implicitly does.  However, most ionizing photons are absorbed within a few Mpc of their host bubble (e.g., \citealt{mcquinn09}); so long as the characteristic bubble size is larger than this, our approximation is reasonable.  We account for the remaining photons via the hard, weak ionizing background described in \S \ref{hard}.

\begin{figure*}
\plottwo{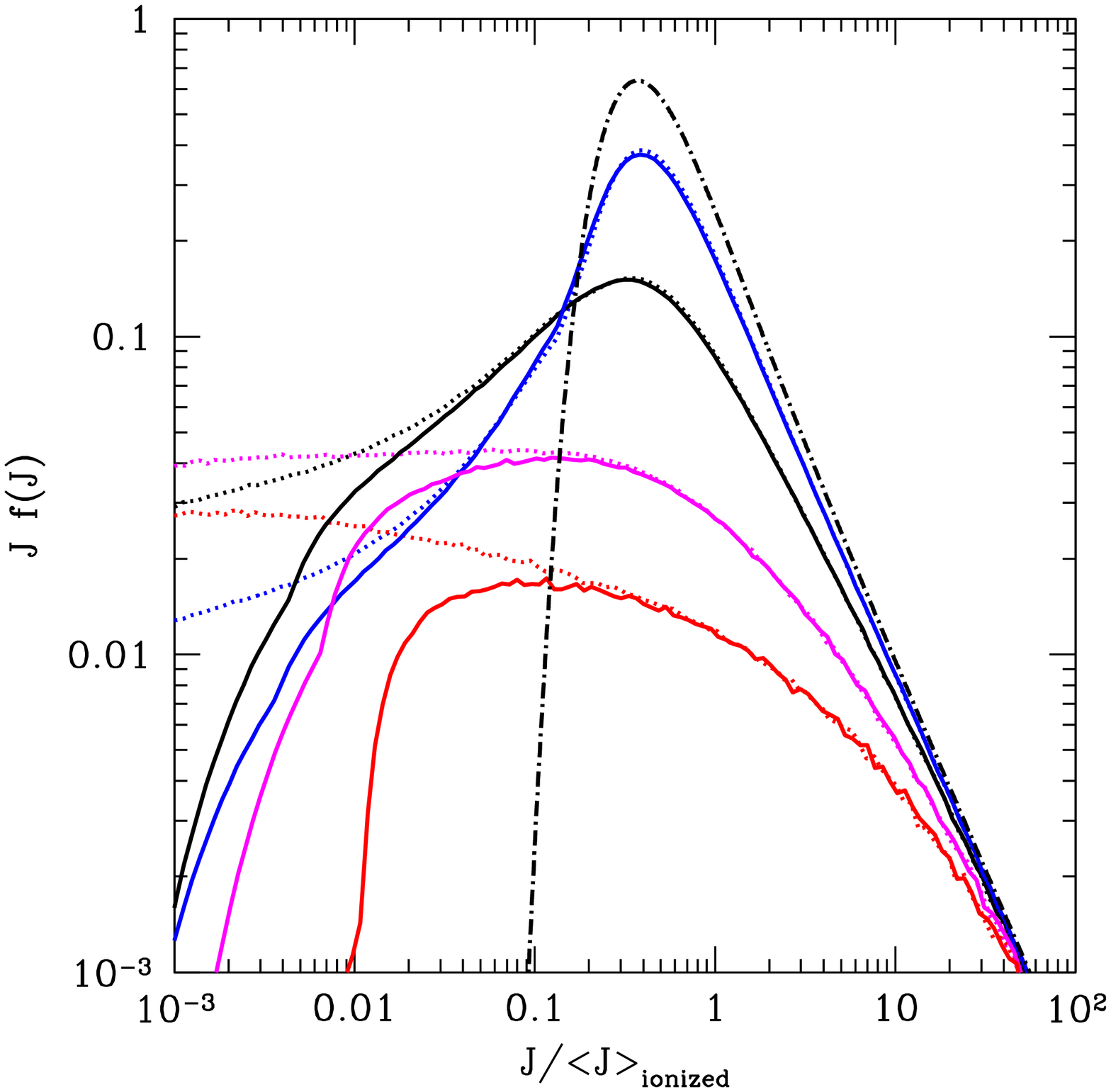}{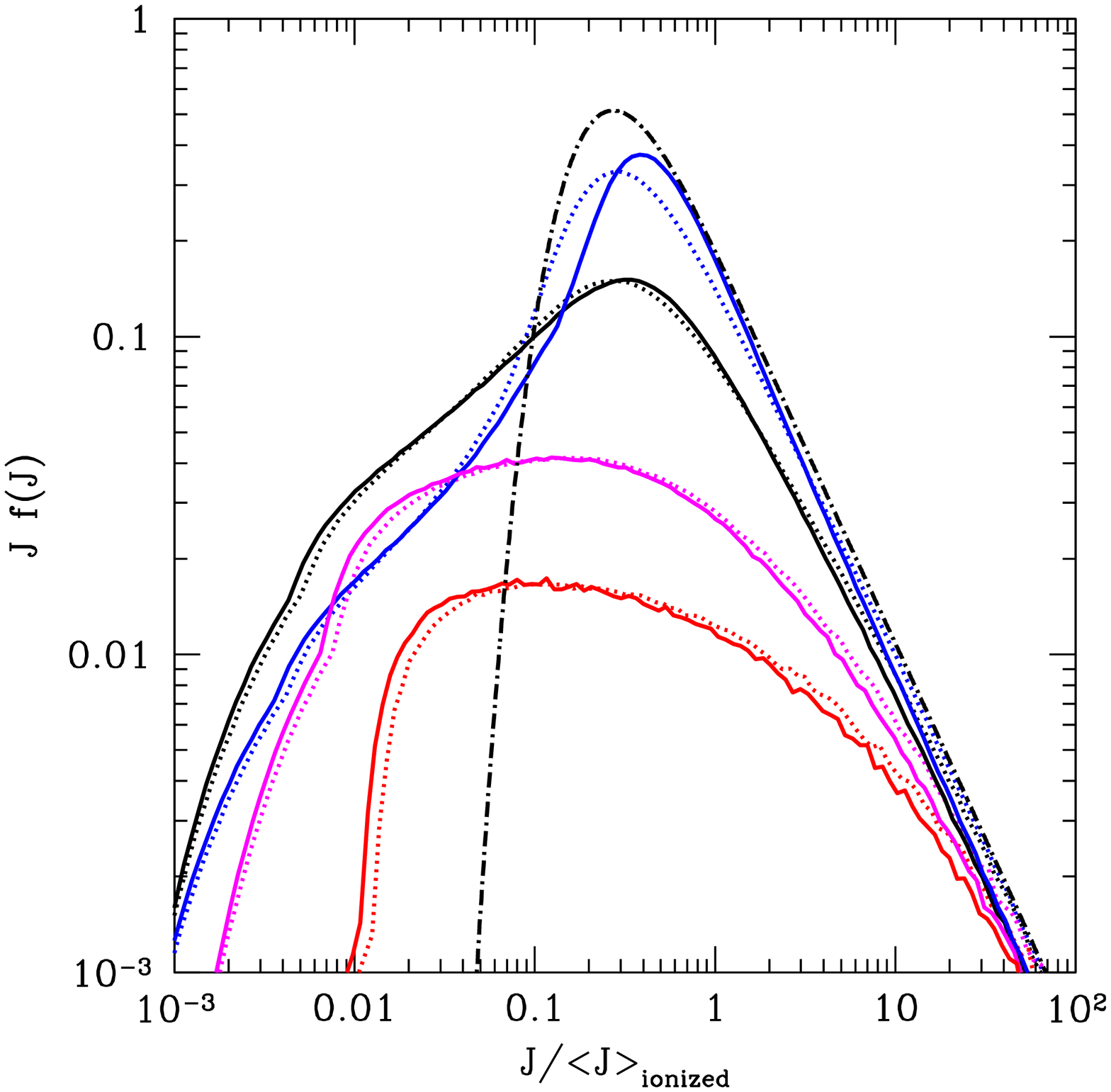}
\caption{Distribution of $J$ during reionization, relative to its mean value in a fully-ionized IGM, $\VEV{J}_{\rm ionized}$.  The solid curves in both panels show $f(J)$ in our fiducial model, with $z=3$, $L_{\rm min}=10^{43} \ergsec$ and $r_0=35 \Mpc$, at $\bar{x}_{\rm HeIII}=0.3,\,0.5,\,0.75,$ and 0.9, from bottom to top at the peak.  The dot-dashed curves show the post-reionization distribution for the same parameters.  \emph{Left panel:}  The dotted curves take $L_{\rm min}=10^{40} \ergsec$, for the same four points during reionization.  \emph{Right panel:}  The dotted curves take $r_0=25 \Mpc$.}
\label{fig:reion_params}
\end{figure*}

Third, the simple ``photon-counting" model we use here does not properly incorporate recombinations, which slow the growth of large bubbles.  These can be included in the excursion-set formalism \citep{furl05-rec, furl08-helium}, but in that case the method returns the distribution of the minimum of $(R,r_0)$.  For simplicity, we take the more transparent route of computing the sizes of discrete regions and then imposing attenuation when calculating the ionizing background.  

Despite these shortcomings, the \citet{furl08-helium} model is still the only concrete calculation of bubble sizes, so we will use it here with the caveat that the distributions are probably not quantitatively accurate.  The crucial aspect is that it has the correct \emph{qualitative} behavior, in which discrete bubbles gradually grow larger and larger, until they surpass the attenuation length (beyond which the total size matters less and less).  Moreover, note that the bubble size distribution at early times is not essential in the limit of rare sources; for example, if each bubble hosts only one source, the ionizing background at interior points is independent of $R$ -- which only determines the minimum allowed $J$.

\subsection{Results}
\label{fj-reion}

Once $n_b(R)$ is specified, we compute the overall distribution of the amplitude of the ionizing background via
\bq
f(J|r_0) = \int dR \, n_b(R) f(J|R,r_0),
\label{eq:fj-reion}
\eq
where we use the Monte Carlo model for $R<90 \Mpc$ and equation~(\ref{eq:jdist_r0}) for larger bubbles.

Figure~\ref{fig:reion_params} shows the resulting distribution in several specific cases.  The solid curves in both panels show our fiducial model, in which $z=3$, $L_{\rm min}=10^{43} \ergsec$ and $r_0=35 \Mpc$.  From bottom to top (at the peak), the curves assume $\bar{x}_{\rm HeIII}=0.3,\,0.5,\,0.75,$ and 0.9.  For comparison, the dot-dashed curve shows the post-reionization distribution for the same parameters.

In general, the trends are similar to those during hydrogen reionization \citep{furl09-mfp}.  Initially, the ionized regions are all relatively small, so any individual point is illuminated by only a few sources and the amplitude of the ionizing background is usually relatively small.  As bubbles grow and encompass more sources, the amplitudes increase.  The rate of increase slows once attenuation becomes important (when the mean bubble size exceeds $r_0$), even though $R$ actually accelerates its growth in this regime \citep{furl08-helium}.  The distributions then match smoothly onto the post-reionization $f(J)$.  The final stage is the incorporation of the remaining small bubbles into large ionized regions, destroying the low-$J$ tail as these previously-isolated points now receive weak illumination from distant quasars.

However, there are important differences with hydrogen reionization.  First, there is less evolution at large $J$ here.  This is partly because the rarity of sources makes even the post-reionization distribution broader and partly because the high-$J$ tail is created by points close to a single quasar, so its amplitude depends only on their abundance -- which is fixed between all these curves.  Thus, the transition from pre- to post-reionization is even smoother than for hydrogen reionization.   

Another feature that differs from hydrogen reionization is the normalization of $f(J)$.  If sources are common enough that every bubble contains at least one, then the integral of $f(J)$ must equal $\bar{x}_{\rm HeIII}$.  However, during helium reionization quasars are sufficiently rare that many smaller bubbles contain no sources.  For example, the mean number of sources per attenuation volume is only 4.2 after reionization is complete (if $r_0=35 \Mpc$ at $z=3$); thus any individual 15 Mpc bubble has only an $\sim 8\%$ chance to contain a quasar (although, during reionization, this increases because quasars can only live inside ionized bubbles).  Thus, when $\bar{x}_{\rm HeIII}=(0.3,\,0.5,\,0.75,\,0.9)$, the actual fraction of space illuminated by ionizing radiation is only $(0.09,\,0.22,\,0.55,\,0.81)$.  This has interesting observable implications; see \S \ref{reion}.

Moreover, we also find that the \emph{mean} $J$ within ionized regions is nearly constant across all these models:  it increase by only $\sim 20\%$ from $\bar{x}_{\rm HeIII}=0.3$--1.  Thus the average $J$, including both \ion{He}{3} regions and the \ion{He}{2} walls between them, is roughly proportional to $\bar{x}_{\rm HeIII}$.  This is not too surprising; we have held the comoving emissivity constant between these models, so $J$ can only increase by allowing photons to propagate farther.  We have $r_0/R_{\rm min} \sim 2.3$, so the mean background can only increase by roughly that factor.  Provided our estimates for the attenuation length are reasonable, helium reionization will therefore \emph{not} be accompanied by an enormous increase in the ionizing background.

The two panels of Figure~\ref{fig:reion_params} also show how these results depend on some of the model's input parameters.  In the left panel, the dotted curves assume $L_{\rm min}=10^{40} \ergsec$.  In this case, ionizing sources are much more common, and $\ga 80\%$ of of ionized regions contain active sources even when $\bar{x}_{\rm HeIII} = 0.3$.  However, these additional sources are all extremely faint.  In reality, the uniform hard photon background (\S \ref{hard}) will provide a lower limit to the ambient ionizing background of $\sim 0.05 \VEV{J}$; Figure~\ref{fig:reion_params} then shows that extremely faint active galaxies will not affect the observable distribution.

The right panel of Figure~\ref{fig:reion_params} shows the distributions if $r_0=25 \Mpc$.  Perhaps surprisingly, this makes very little difference to $f(J)$.  One reason is that we have scaled to the mean value after reionization, which is proportional to $r_0$; for a fixed quasar emissivity, $\VEV{J}_{\rm ionized}$ \emph{will} increase with the attenuation length.  However, the shapes of the distributions are nearly invariant, which can be understood by referring to Figure~\ref{fig:mw_ex} and by noting that attenuation has only a minimal effect on small bubbles.

\begin{figure}
\plotone{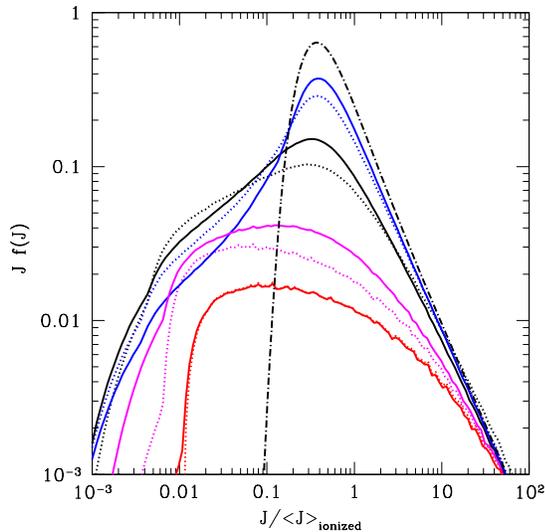}
\caption{Distribution of $J$ during reionization, relative to its mean value in a fully-ionized IGM, $\VEV{J}_{\rm ionized}$.  The solid curves show $f(J)$ in our fiducial model, with $z=3$, $L_{\rm min}=10^{43} \ergsec$, and $r_0=35 \Mpc$, at $\bar{x}_{\rm HeIII}=0.3,\,0.5,\,0.75,$ and 0.9, respectively.  The dot-dashed curves show the post-reionization distribution for the same parameters.  The dotted curves show the same four phases during reionization, but allowing quasars to sit inside smaller halos (see text).}
\label{fig:reion_mlow}
\end{figure}

Figure~\ref{fig:reion_mlow} shows how $f(J)$ depends on $n_b(R)$.  The solid curves again show our fiducial model.  The dotted curves are identical, except they assume that quasars sit inside halos with $m \ga 5 \times 10^{10} \Msun$, an order of magnitude smaller than the fiducial model.\footnote{This does not affect the number density of active quasars -- which  need not sit inside every such halo.}  These halos are less strongly clustered, which shifts $n_b(R)$ toward somewhat smaller radii.  The net effect is that more of the Universe sits inside of small bubbles, so $f(J)$ shifts leftward (and its normalization decreases, because bubbles are more likely to be empty).  Evidently the details of the bubble size distribution do not affect the qualitative trends of the model.

We should note that we have assumed a fixed redshift ($z=3$) and attenuation length in these figures.  Of course, reionization will actually occur over a finite time interval (although a small one, if current estimates hold; \citealt{furl08-helium, mcquinn09}), over which the luminosity function and the attenuation length itself may evolve.  Figure~\ref{fig:mw_ex} shows that evolution in the luminosity function has only a minimal effect.  Moreover, Figure~\ref{fig:rmax} shows that attenuation around individual quasars -- the limit most relevant during reionization -- does not evolve either.  On the other hand, once a more uniform background builds up (at $\bar{x}_{\rm HeIII} \ga 0.8$, according to Fig.~\ref{fig:reion_params}), estimates based on the \ion{H}{1} \lya forest predict $r_0 \propto (1+z)^{-3}$ or so \citep{madau99-qso, faucher08-ionbkgd}.  This may accelerate the evolution during the final phases of reionization, although the short available time interval makes the total change small -- so we still securely expect \emph{smooth} evolution of the ionizing background onto its post-reionization form, rather than a sharp jump (c.f. \citealt{furl09-mfp} for \ion{H}{1} reionization).  

\subsection{$\eta$ During Reionization}
\label{reion}

The hardness parameter $\eta$ provides one way to constrain this smooth evolution.  As with our post-reionization distributions, we can easily transform $f(J)$ into $f(\eta)$ during reionization, assuming that all points are in ionization equilibrium (see below).  Figure~\ref{fig:eta_reion} shows the results.  The upper and lower panels show the probability distribution and the cumulative distribution function, respectively.  In each panel, the dot-dashed curve shows our fiducial post-reionization model, with $z=3$, $L_{\rm min}=10^{43} \ergsec$, and $r_0=35 \Mpc$.  The solid curves take the same parameters but assume $\bar{x}_{\rm HeIII}=0.3,\,0.5,\,0.75,$ and $0.9,$ from bottom to top.  Again, we have normalized these so that, in the post-reionization Universe, $\VEV{\eta}=60$.

\begin{figure}
\plotone{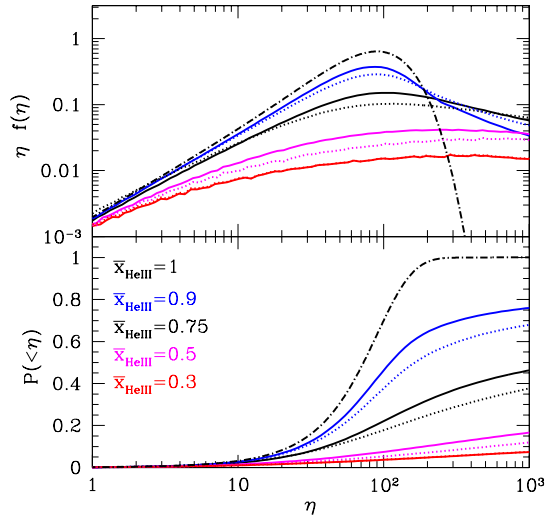}
\caption{Distribution of the hardness ratio, $\eta$, during helium reionization.  The solid curves show our fiducial model, with $z=3$, $L_{\rm min}=10^{43} \ergsec$, and $r_0=35 \Mpc$, at $\bar{x}_{\rm HeIII}=0.3,\,0.5,\,0.75,$ and $0.9,$ from bottom to top.  The dotted curves are same, except they allow quasars to live in smaller dark matter halos; note that the $\bar{x}_{\rm HeIII}=0.3$ curves overlap in both panels.  The dot-dashed curve shows the post-reionization distribution, for the same parameters.}
\label{fig:eta_reion}
\end{figure}

The distributions during reionization are markedly different from those after the process has completed, principally because a large fraction of the points have $\eta \gg 100$ -- and indeed those either entirely outside of ionized bubbles or inside of an empty one have an essentially infinite value (see \S \ref{fossil} for a discussion of the latter).  Interestingly, the fraction of such pixels is significantly \emph{larger} than $\bar{x}_{\rm HeII}$, because of the many points inside empty (but mostly ionized) regions.  This exaggerates the ``apparent" \ion{He}{2} fraction that one might naively associate with the fraction of pixels with $\eta \gg 100$ and appears to make helium reionization a sharply defined, dramatic event with this observable (even though, as we have emphasized above, $f(J)$ does match smoothly onto the post-reionization form, and in a regular fashion).

This smoothness does manifest itself in $f(\eta)$ as the relative stability of the low-$\eta$ tail.  Our models predict that, at least during the latter half of reionization, a substantial fraction of the Universe will have $\eta \la 100$:  in other words, \emph{regions of high transmission will continue to appear well into the helium reionization era}.  This is a key difference from hydrogen reionization, where the overall saturation of \ion{H}{1} absorption makes studies of the \lya forest almost impossible before the hydrogen becomes fully ionized.  Note that these regions do \emph{not} need to be extraordinarily close to bright quasars; they simply need the local background to be comparable to the post-reionization value, or to have a few $L_\star$ quasars within a couple of attenuation lengths.\footnote{One possible complication is the damping wing of \ion{He}{2} outside of the ionized region, which is extremely important during \ion{H}{1} reionization \citep{miralda98}.  However, the optical depth of \ion{He}{2} is two orders of magnitude smaller, so the wing is much weaker during this time and will not interfere with most of the large ionized bubbles that we expect.}

Thus the differing character of helium reionization -- and especially the more rapid buildup of large ionized regions and the dominance of rare bright sources -- suggests that spectral studies of that reionization era will continue to provide a great deal of information about the IGM; we do \emph{not} by any means expect complete saturation in the forest at $z \ga 3$.  Followup of the many potential \ion{He}{2} \lya forest lines of sight recently discovered through the Sloan Digital Sky Survey and the GALEX mission should be extremely fruitful \citep{zheng04-sdss, zheng08, syphers08}.

Recall that the ``reliable" $\eta$ distribution from \citet{fechner06} showed a longer tail toward high $\eta$ than permitted in our post-reionization distributions.  This could, in principle, indicate that this line of sight is probing the patchy phase of \ion{He}{2} reionization, where small ionized bubbles give much bigger values of $\eta$.  However, we would expect a very long tail if  this were the case, rather than simply extending slightly farther than the post-reionization models.  We therefore regard this interpretation as unlikely at this point, although if it were the case then this would indicate that \ion{He}{2} reionization did not complete until $z \la 2.7$.

The dotted curves in each panel show the distributions for a model in which quasars are allowed to inhabit smaller dark matter halos (as in Fig.~\ref{fig:reion_mlow}), so that the ionized bubbles are also smaller.  (Note that the $\bar{x}_{\rm HeIII}=0.3$ curve overlap in both panels.)  In this case, they are more likely to lack active sources, so the curves lie below the fiducial model.  In principle, this can be used to measure (crudely) the characteristic bubble size at a given stage of reionization, although at the moment it is probably more accurate to view the difference as the uncertainty in simple models.  Detailed interpretations of the data will require careful comparison to simulations, such as those of \citet{mcquinn09} or \citet{paschos07}.  This is particularly important toward the end of reionization, when the growing hard photon background may help to eliminate the last vestiges of \ion{He}{2}, making the evolution during these last phases smoother than in our simple models.

Simulations are also necessary to quantify the complex effects of radiative transfer during \ion{He}{2} reionization, which may broaden the distribution substantially \citep{tittley07}, including especially a wider tail toward high values of $\eta$.  We have also ignored variations in $\alpha$, which introduce additional fluctuations in the frequency-integrated $\Gamma$.

We have neglected the nearly uniform, optically thin background in Figure~\ref{fig:eta_reion}, but it provides $J \sim 0.05 \VEV{J}$ everywhere.  At this level, $\eta \sim 1200$, off the right edge of our plot.  Measuring the growth of this weak background will probably be extremely difficult.  If the \lya forest lines are broadened only by turbulence or peculiar motions, the hardness ratio can be related to the optical depths of pixels in the \ion{H}{1} and \ion{He}{2} forests through $\eta = 4 \tau_{\rm HeII}/\tau_{\rm HI}$ \citep{miralda93}.  In practice, reliable measurements require $0.01 \la \tau \la 1$, which limits the total measurable range of the hardness ratio to $\eta \sim 1$--$400$.  Thus, the high-energy background is probably not measurable with line ratios.

\subsection{Fossil Bubbles}
\label{fossil}

We now re-visit the properties of ionized bubbles without any active sources.  Above we assumed that they have $\eta \gg \VEV{\eta}$ and were essentially invisible -- even though they still may be highly ionized overall.  These regions initially sat in photoionization equilibrium with their original ionizing source.  Once that source shut off, they fall out of equilibrium (in the sense of being over-ionized) and become ``fossil" bubbles that recombine rapidly -- although also non-uniformly -- thanks to the clumpiness of the intergalactic medium \citep{furl08-fossil}.  The ratio of the helium recombination time to the Hubble time is
\bq
{t_{\rm rec} \over H^{-1}(z)} \approx {0.4 \over \Delta} \left( {4 \over 1+z} \right)^{3/2} \left[ {\alpha_A(2 \times 10^4 \kel) \over \alpha} \right],
\label{eq:rectime}
\eq
where we have assumed matter domination and a slightly higher temperature than before (because the region was recently ionized; \citealt{furl08-igmtemp, mcquinn09}).  Thus these regions will remain relatively highly ionized for of order the expansion timescale -- particularly in the low-density voids which provide most of the transmission in the \ion{He}{2} \lya forest.

However, strong absorption in \ion{He}{2} requires only a small fraction of the material to have recombined (see eq.~\ref{eq:taugp}).  How rapidly will a typical feature become saturated in absorption?  For a rough estimate, we let $x_{\rm HeII} \approx \alpha(T) n_e(\Delta) t_{\rm fossil}$ (valid when $x_{\rm HeII} \ll 1$) and find
\bq
\tau_{\rm HeII} \approx 180 \Delta^2 \left( {t_{\rm fossil} \over 5 \times 10^7 \yr} \right) \left( {1+z \over 4 } \right)^{9/2}.
\label{eq:crit-recomb}
\eq
Thus, all regions with $\Delta \ga 0.1$ will already be saturated \ion{He}{2} absorbers just 50 million years after the quasar shuts off.  In the simple \lya forest model of \citet{schaye01}, this includes all \lya forest systems with \ion{H}{1} column densities $\ga 10^{12} \cmden$ (or $\tau_{\rm HI} \ga 0.01$) at $z \sim 3$, and in simulations such systems cover $\ga 95\%$ of the volume of the Universe \citep{miralda00}.  Thus the vast majority of the observable IGM will recombine quickly after their illuminating source expires.

The recombining regions will provide a smoother transition to high $\eta$ values, but they will typically be very difficult to observe.  Only rare voids can still provide transmission over long timescales.  (Of course, all these regions will eventually equilibrate with the uniform, high-energy background of \S \ref{hard}.  But, as noted previously, this background is not strong enough to keep absorbers transparent.)

\section{Discussion}
\label{disc}

We have examined the evolution of the high-energy ionizing background during and after helium reionization.  Through a combination of analytic calculations and a Monte Carlo model, we have computed the probability distribution of $J$, the amplitude of the ionizing background at the ionization edge (which is a reasonable proxy for the ionization rate, $\Gamma$).  After reionization is over, our model requires only the attenuation length of helium-ionizing photons and the quasar luminosity function (in the relevant energy range) as inputs.  We find that bright quasars remain rare enough at $z \sim 2$--3 that $f(J)$ is quite broad, with a full-width-at-half-maximum comparable to $\VEV{J}$, in agreement with \citet{meiksin07}.  

Currently, the distribution of the hardness ratio $\eta = N_{\rm HeII}/N_{\rm HI}$ is the best way to measure such fluctuations \citep{zheng04, shull04, fechner06, fechner07}.  The best-observed distribution \citet{fechner06} is slightly broader than our models predict, especially toward high $\eta$, but the overall agreement is good (see also \citealt{meiksin07}).  Some of this additional broadening is undoubtedly due to more complex radiative transfer than we include here as well as variations in the spectral indices of the ionizing sources.  Note that there also appear to be observational biases affecting many of the estimates in the literature.  For example, the apparent optical depth method becomes unreliable in saturated lines \citep{fechner06, fechner07}.  A more careful comparison of detailed theoretical models and the data is still needed in order to quantify the tension.

During reionization, quasars sit inside of discrete ionized bubbles; the ``walls" between them are full of \ion{He}{2} that blocks low-energy photons from more distant sources.  Thus $f(J)$ depends on the size distribution of these bubbles.  With a simple analytic model of this distribution, we have used a Monte Carlo method (that also includes attenuation) to generate $f(J)$ in this regime.  Initially, $f(J)$ is quite broad because of the range of bubble sizes (and because most bubbles are sufficiently small that a faint quasar can provide all of the local ionizing radiation field).  As the bubbles grow larger, they encompass more and more sources and the variance slowly declines.  Eventually, the distribution matches smoothly onto its post-reionization form:  after $\bar{x}_{\rm HeIII} \sim 0.9$, the only significant change is the disappearance of the low-$J$ tail from isolated regions.  Thus we do \emph{not} expect a sharp feature in the overall amplitude of the ionizing background at the completion of reionization.

We found that the mean ionizing background within illuminated regions remains roughly constant.  This contrasts with the situation during hydrogen reionization, where the same quantity increases in proportion to the characteristic bubble size.  The difference occurs because so few sources contribute to helium reionization:  only a few are needed to reach the mean post-reionization value inside a bubble.  During hydrogen reionization, on the other hand, the number of visible sources increases proportionally to the volume of the bubble.

Despite this smooth match onto the post-reionization Universe, there are fairly sharp observational probes of the reionization era.  Most importantly, we have shown that the fraction of pixels with large values of $\eta$ ($\ga 1000$) increases faster than linearly as $\bar{x}_{\rm HeII}$ increases.  Not only is the hardness ratio large inside of regions that have not yet been fully-ionized, but ionized bubbles without any active sources also rapidly recombine and become nearly opaque in \ion{He}{2}.  This should make \ion{He}{2} reionization \emph{easier} to identify.  

This contrasts with the post-reionization Universe, where $\eta \la 200$ everywhere in our model; this maximum value corresponds to the minimum ionizing background generated by distant quasars.  Our model shows that this should be a rather sharp cutoff (at least without shadowing and radiative transfer; \citealt{tittley07}), because the probability of having zero bright sources within two attenuation lengths is very small.  Direct comparisons to data are difficult, however, because real observations inevitably contain a number of lower limits to $\eta$ (corresponding to points where the \ion{H}{1} optical depth is too small to measure).  A careful calibration to more detailed simulations is required to draw firm conclusions about the timing of helium reionization from the $\eta$ distributions.

On the other hand, the high-$J$ tail of the distribution (corresponding to small $\eta$) -- which easily reaches values several times the mean -- remains more or less intact throughout at least the latter half of reionization.  Thus, we expect that measurable pockets of transmission in the \ion{He}{2} \lya forest \emph{will} persist into the reionization epoch at $z \ga 3$:  a true \citet{gunn65} trough will only appear when nearly all of the helium atoms are singly ionized.  This implies that ultraviolet spectra of higher-redshift quasars will continue to provide detailed information about this important transformational epoch, even at redshifts well beyond the nominal ``end" of reionization.  Fortunately, dozens of promising targets have now been detected \citep{zheng04-sdss, zheng08, syphers08}, and the recent installation of the Cosmic Origins Spectrograph and repair of the Space Telescope Imaging Spectrograph on the \emph{Hubble Space Telescope} has provided two powerful instruments for such exploration.

Note that this also contrasts with hydrogen reionization, where variations in the ionizing background are small during the final phases \citep{furl09-mfp}.  There, the huge number of galaxies responsible for reionization markedly decreases the importance of the high-$J$ tail.  Without such strongly-ionized regions, and with the high mean densities at these redshifts, the \ion{H}{1} \lya forest becomes completely saturated even inside of large ionized bubbles before reionization completes; the appearance of a \citet{gunn65} trough at $z \sim 6$ may or may not hint to us that reionization is ending at that time, but in either case the prospects for detecting substantial transmission at higher redshifts are dim.  That is not the case for helium, for which our model strongly predicts that regions of transmission will be relatively common throughout the bulk of that process.

Careful study of the \ion{He}{2} forest during the reionization era may help to illuminate some of the physics of that time.  For example, comparing the hydrogen and helium absorption lines may help us to identify when thermal broadening is important to the forest (and hence when, and where, helium reionization photoheats the IGM; \citealt{gleser05, furl08-igmtemp, mcquinn09, bolton08}).  The spatial scales over which the hardness ratio fluctuates will also constrain the growth of \ion{He}{3} bubbles, and of the attenuation length.  These will no doubt be difficult measurements, but they will provide more direct information about the equation of state of the IGM than any other method.

Our model does not, of course, provide a complete picture of the ionizing background during helium reionization.  Most importantly, we only consider the effects of radiative transfer in the crudest possible manner (via a fixed attenuation length $r_0$).  Shadowing, fluctuations in the mean free path, and the large range of relevant frequencies and spectral indices will all modify the simple picture here \citep{shull04, maselli05, tittley07}.  Another aspect that we ignore is quasar clustering, which is fairly strong \citep{shen07, francke08}, although still probably sub-dominant to random fluctuations \citep{mcquinn09}.  

Another important detail that our model does not directly describe is the evolution of $r_0$.  We have argued that the attenuation length as seen by a single quasar does not evolve significantly (Fig.~\ref{fig:rmax}); however, once a universal, more spatially constant background has built up, $r_0$ will converge to a more uniform value.  Reassuringly, these two approaches give similar mean free paths at $z \approx 3$ (see also \citealt{furl08-helium}), roughly where they must converge to each other at the end of reionization.  On the other hand, in our figures we have kept redshift constant for a straightforward comparison; if $r_0$ evolves rapidly there will be more evolution in $f(J)$ than we have shown here (although this is only important once the characteristic bubble size exceeds $r_0$).

In this paper, we have focused on how fluctuations in the helium-ionizing background affect the hardness ratio of forest lines.  Of course, the wide variation in $J$ even after helium reionization ends may be observable in other ways.  For example, it will affect the evolution of the mean optical depth in the \ion{He}{2} forest:  strongly illuminated regions will allow substantial transmission in some regions even when most of the Universe is opaque (see, e.g., \citealt{furl09-mfp} for a similar treatment for hydrogen reionization).  This will cause a more gradual evolution in the overall optical depth $\tau_{\rm eff}$ than predicted by naive reionization models.  Moreover, the high-energy radiation field can also be measured with metal lines.  The data so far are controversial:  some measurements imply a hardening in the radiation field at $z \sim 3$ \citep{songaila98, songaila05, agafonova07}, while others are consistent with no evolution \citep{kim02, aguirre04}.  The complexities of radiative transfer during helium reionization may account for these \citep{madau08}.  Alternatively, the strongly fluctuating ionizing background throughout reionization provides a natural explanation for these differences, although of course a quantitative estimate is still required.

\acknowledgments

I thank the anonymous referee, A.~Mesinger, and J.~Bolton for helpful comments on the manuscript, and J.~Bolton, C.~Fechner, J.~M.~Shull, and J.~Tumlinson for sharing their data in electronic form.  This research was partially supported by the NSF through grant AST-0829737 and by the David and Lucile Packard Foundation.

\bibliographystyle{apj}
\bibliography{Ref_composite}

\end{document}